\documentstyle[12pt,times]{article}
\newlength{\abstractwidth}
\renewcommand{\thefootnote}{\fnsymbol{footnote}}
\renewcommand{\thanks}[1]{\footnote{#1}} 
\newcommand{\starttext}{
\setcounter{footnote}{0}
\renewcommand{\thefootnote}{\arabic{footnote}}}

\newcommand{\be}{\begin{equation}}
\newcommand{\bea}{\begin{eqnarray}}
\newcommand{\eea}{\end{eqnarray}}
\newcommand{\beq}{\begin{equation}}
\newcommand{\ee}{\end{equation}}
\newcommand{\eeq}{\end{equation}}

\def\ba{\begin{eqnarray}}
\def\ea{\end{eqnarray}}
\newcommand{\PSbox}[3]{\mbox{\rule{0in}{#3}\includegraphics{#1}\hspace{#2}}}

\def\half{{1 \over 2}}
\def\d{{d \over 2}}

\begin{document}
\begin{titlepage}
\bigskip
\hskip 3.7in\vbox{\baselineskip12pt
\hbox{MIT--CTP--2857}
\hbox{UCLA/99/TEP/16}
\hbox{hep-th/9905049}}
\bigskip\bigskip\bigskip\bigskip

\centerline{\Large \bf AdS/CFT 4--point functions:}
\smallskip
\centerline{\Large \bf How to succeed at $z$--integrals without really
trying}

\bigskip\bigskip
\bigskip\bigskip

\centerline{ Eric D'Hoker$^{a}$, Daniel Z. Freedman$^{b,c}$, 
and Leonardo Rastelli$^{b,}$\footnote[1]{\tt dhoker@physics.ucla.edu, dzf@math.mit.edu, rastelli@ctp.mit.edu.}}
\bigskip
\bigskip
\centerline{$^a$ \it Department of Physics}
\centerline{ \it University of California, Los Angeles, CA 90095}
\centerline{\it and  Institute for Theoretical Physics}
\centerline{\it University of California, Santa Barbara, CA 93106}
\bigskip
\centerline{$^b$ \it Center for Theoretical Physics}
\centerline{ \it Massachusetts Institute of Technology}
\centerline{ \it Cambridge, {\rm MA}  02139}
\bigskip
\centerline{$^c$ \it Department of Mathematics}
\centerline{ \it Massachusetts Institute of Technology}
\centerline{\it Cambridge, {\rm MA} 02139}
\bigskip\bigskip

\begin{abstract}
A new method is  discussed which vastly simplifies one of the two 
integrals over $AdS_{d+1}$ required to compute exchange graphs
for 4--point functions of
scalars in the AdS/CFT correspondence. The explicit form
of the bulk--to--bulk propagator is not required. Previous results for
scalar, gauge boson and graviton exchange are reproduced, and new results 
are given for massive vectors.
It is found that precisely for the cases that occur in the
$AdS_5 \times S_5$ compactification of Type IIB supergravity,
the exchange diagrams reduce to a {\it finite} sum of
graphs with quartic scalar vertices.
The analogous integrals
in $n$--point scalar diagrams for $n>4$ are also evaluated.

\end{abstract}

\end{titlepage} 
\starttext
\baselineskip=18pt
\setcounter{footnote}{0}
\section{Introduction }
During the past year several groups have calculated 4--point correlation 
functions in AdS supergravity as part of the study of the AdS/CFT
correspondence \cite{maldacena,polyakov, witten}. 
In particular the position space correlators for 
quartic scalar
interactions \cite{canadians, august} , gauge boson exchange \cite{dhfgauge},
 scalar field exchange \cite{liu, dhfscalar},
and graviton exchange \cite{march99} have been obtained. 
There is additional work on a momentum space approach \cite{chsch}.

Exchange diagrams, see Figure 1, contain a bulk--to--bulk propagator, and two
integrations over $AdS_{d+1}$ are required to compute the amplitude. 
In past work the first integral, called the $z$--integral, was calculated by a 
cumbersome expansion and resummation procedure which
typically  gave a
 simple function of the other bulk coordinate $w_\mu$ as result. This suggests
that a more direct method should be possible, and it is the main purpose of the
present paper to present one. Specifically we show that $z$--integrals satisfy
a simple differential equation which can be solved recursively.
The specific form of the bulk--to--bulk propagator is not required. 
All previous
cases can be handled quite easily by the new method, and we are also
able to obtain new results for massive vector 
exchange amplitudes as well as for higher point correlators.
The new method does not simplify the remaining integral over the $w_\mu$
coordinate, and we refer to past work \cite{dhfgauge,dhfscalar,march99} 
in which useful
integral representations and asymptotic formulas for these $w$--integrals
have been derived.

Our main focus of interest is the 
$AdS_5 \times S_5$ compactification of IIB supergravity \cite{vannieuv}, 
but clearly the method we propose, and most of our formulas,
have a general validity. We do not  discuss
certain subtleties that occur in $d=2$ for massless vector and 
graviton equations,
which would require  a more careful investigation
of asymptotics and are
left to future work (hopefully by someone else).

In all the exchange graphs that we study,
it is found that precisely for the trilinear couplings
that occur in the $AdS_5 \times S_5$ supergravity, the
exchange diagram reduces to a {\it finite} sum of scalar
quartic graphs. Generic couplings give instead
an infinite sum. We lack a fundamental
explanation of this fact, although we suspect
some simple mathematical reason
related to harmonic analysis on $S_5$ and representation
theory of the conformal group $SO(5,1)$.
It would be interesting
 to check whether the same holds for other
supergravity compactifications of interest in the AdS/CFT
correspondence (for example \cite{compactifications}).

The basic idea is presented in Section 2 for scalar exchange. 
Massless and massive
vector exchange is discussed in Section 3, and graviton exchange in Section 4.
In Section 5 we discuss an application to $n$--point correlators for $ n \geq 5$.


\begin{figure} 
\begin{center} \PSbox{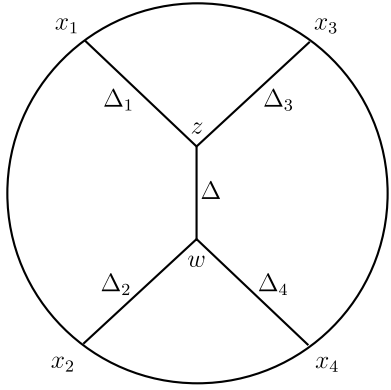 
hscale=80 vscale=80}{7.1in}{5.4in} \end{center} \vspace{-8cm}
\begin{center} \caption{A general t--channel exchange diagram. 
 } \end{center}
\end{figure}


\section{Scalar exchange}
As in most past work, we calculate on the Euclidean continuation of $AdS_{d+1}$,
which is modelled as the upper half space  $z_\mu \in {\bf R}^{d+1}$, 
with $z_0 >0$, and metric $g_{\mu \nu}$ of
constant negative curvature $R=-d(d+1)$, given by
\be
  ds^2 = \sum _{\mu, \nu=0} ^ d g_{\mu \nu} dz_\mu dz_\nu
       ={1 \over z_0^2} (dz^2_0 + \sum ^d_{i=1} \, dz^2_i)
\, .
\ee
The Christoffel symbols are
\be
\Gamma ^\kappa _{\mu \nu} 
 = {1 \over{ z_0}} \bigl ( 
 \delta _0 ^\kappa \delta_{\mu \nu}- \delta _{\mu 0} \delta ^\kappa _\nu 
                   - \delta _{\nu 0} \delta ^\kappa _\mu \bigr )\,.
\ee
It is well-known that AdS--invariant functions, such as scalar propagators, are
simply expressed \cite{prop}
 as functions of the  chordal distance  $u$, defined by
\be \label{u}
  u  = {(z-w)^2 \over 2z_0 w_0}
\qquad \qquad
(z-w)^2 = \delta_{\mu \nu}(z-w)_{\mu} (z-w)_{\nu}
\ee
A scalar field of mass $m^2$ is characterized by two possible scale dimensions,
namely the roots
\be
\Delta_{\pm} = \frac{d}{2}  \pm \frac{1}{2}\sqrt{d^2 + 4m^2}
\ee
of the quadratic relation $m^2 = \Delta(\Delta-d)$. The mass must satisfy
the bound \cite{bfbound}
 $m^2 \geq  -d^2/4$. For $m^2 \geq -d^2/4 + 1$, one must choose
the largest root $\Delta=\Delta_+$. In the range $-d^2/4 < m^2 < -d^2/4 + 1$,
the bulk field may be quantized with either dimension $\Delta_{\pm}$, and it
is known 
 that supersymmetry can require both choices to occur in the
same theory. Only the largest root appears in most applications of the 
AdS/CFT correspondence, but we will need to discuss the other possibility
briefly. Unless explicitly indicated $\Delta$ will mean $\Delta_+$. 

The scalar bulk--to--bulk propagator for dimension $\Delta=\Delta_{\pm}$ 
was obtained in \cite{prop}, 
\bea
G_\Delta (u) \label{scalarprophyper}
& =&
  \tilde C _\Delta (2u^{-1} )^\Delta 
  F \bigl (\Delta , \Delta -\d +\half; 2\Delta -d +1; -2u^{-1} \bigr ) \\  
\tilde C _\Delta & =& {\Gamma (\Delta) \Gamma (\Delta -\d +\half)
                        \over  (4\pi)^{(d+1)/2} \Gamma (2 \Delta -d+1)}
\eea
where $F$ is the standard hypergeometric function ${}_2F _1$. The propagator
satisfies the  differential equation
\be \label{scalardelta}
( -\Box  + m^2)G_\Delta (u) = \delta(z,w)
\ee

The scalar bulk--to--boundary propagator for dimension $\Delta$ is given by 
\cite{witten}
\be
   K_\Delta (z, \vec{x}) 
       = \left( 
                 {z_0 \over z^2_0 + \left(\vec{z} - \vec{x} \right)^2}
                  \right)^\Delta \, ,
\ee
where $\vec{x}$ indicates a point on the $d$--dimensional 
boundary of $AdS_{d+1}$.
In this paper we systematically omit the normalization factors 
for bulk--to--boundary propagators \cite{april},
\be
C_\Delta =  {\Gamma (\Delta) \over \pi^{d/2} \Gamma (\Delta - d/2)} \,.
\ee

The integrals we have to evaluate take the form
\be \label{scalarw}
S(\vec x_1, \vec x_2,\vec x_3,\vec x_4)=\int { d^{d+1}w \over w^{d+1}_0}\,
A(w,\vec{x_1},\vec{x_3})\,
 K_{\Delta_2} (w, \vec x_2 )  K_{\Delta_4} (w, \vec x_4 )\,,
\ee
where
\be \label{scalarz}
A(w,\vec{x_1},\vec{x_3}) = \int { d^{d+1}z \over z^{d+1}_0}
 G_\Delta (u)  
  K_{\Delta_1} (z, \vec x_1 )  K_{\Delta_3} (z, \vec x_3 )\,.
\ee
All scaling dimensions will be always understood to be
$\geq \frac{d}{2}$. More general integrals
with derivative couplings can be reduced to this case
(see for example (A.5) in \cite{march99}). In this paper
we develop a new method to calculate the $z$--integrals
(\ref{scalarz}). The remaining $w$--integral (\ref{scalarw})
can then be handled by the asymptotic expansion techniques
developed in \cite{dhfgauge,dhfscalar,march99}.

As in past work, the integral (\ref{scalarz}) is considerably simplified by performing the
translation $\vec x_1 \to 0$, $\vec x_3 \to \vec x_{31} \equiv \vec x_3 - \vec x_1$
and the  
conformal inversion 
\be \label{inversion}
\vec x_{13} =  \frac{ \vec x'_{13}}{|\vec x'_{13}|^2} \qquad
z_\mu = \frac {z'_\mu}{(z')^2} \qquad w_\mu = \frac {w'_\mu}{(w')^2}\,.
\ee
The integral takes the form
\be \label{scalarform}
A(w,\vec x_1,\vec x_3) = | \vec x_{13}|^{-2\Delta_3}\,
 I(w'-\vec x'_{13})
\ee
where
\be \label{scalarintegral}
I(w) = \int {d^{d+1}z \over z^{d+1}_0} \, G_\Delta (u) \,(z_0)^{\Delta_1}
           \left( \frac{z_0}{z^2} \right)^{\Delta_3}\,.
\ee
Convergence of the integral requires $\Delta >|\Delta_1-\Delta_3|$, and we
assume that this condition, and the previous conditions $\Delta, \Delta_i 
\geq \frac{d}{2}$ hold in the following.
Integrals of this form with scalar, vector, and symmetric tensor 
bulk--to--bulk propagators are the main focus of this paper.

Let us first discuss briefly the old method for evaluating the integral
and then the new one. In the old method a quadratic transformation of the
hypergeometric function, namely 
 \be
G_\Delta (u) 
 =
   2^\Delta \tilde C _\Delta \xi ^\Delta 
   F \big (\frac{\Delta}{2}, \frac{\Delta}{2} + \half; \Delta -\d +1;\xi ^2 \big )
\,
\ee
with variable
\be
\xi \equiv {1 \over 1+u} = { 2 z_0 w_0 \over z^2_0 + w^2_0 + (\vec{z} -
  \vec{w})^2 } \, 
\ee
was used.  The propagator was then expressed as a power series in $\xi$ and
the $z$--integral was done term by term using Feynman parameters. The
resulting series, usually a geometric series, was then resummed. For
favorable relations among the dimensions $\Delta$, $\Delta_1$, $\Delta_3$, and 
$d$, relations which cover all the cases in the application of the AdS/CFT
correspondence to the $d=4$, ${\cal N}=4$ superymmetric Yang--Mills theory, the
Feynman parameter integral could also be done and the result for $I(w)$
was a simple polynomial in the variable $w_0^2/w^2$. 

The new method is ultra--simple for scalar exchange. We first note that
invariance of $I(w)$ under the scale transformation $w_\mu \to
\lambda w_\mu$ and under the $d$--dimensional Poincare subgroup of
$SO(d+1,1)$ implies that $I(w)$ can be represented as
\be \label{If}
I(w)= (w_0)^{\Delta_{13}} f(t)
\ee
 where
\be \label{t}
 t= \frac{w_0^2}{w^2} =\frac{w_0^2}{w_0^2+|\vec w|^2}
\ee
 and $\Delta_{13} \equiv \Delta_1-
\Delta_3$. Next we apply the wave operator $(-\Box + m^2)$ to $I(w)$
and use (\ref{scalardelta}) to obtain
\be
(-\Box  + m^2)[(w_0)^{\Delta_{13}} f(t) ] = (w_0)^{\Delta_{13}}
t^{\Delta_3} \,.
\ee
The next step is to work out the action of the
Laplacian on the left side,
which leads to the
inhomogeneous
second order differential equation for $f(t)$
\be \label{scalardiffequ}
4t^2(t-1) f'' + 4t [(\Delta_{13} +1)t -\Delta_{13}+\frac{d}{2}-1]f'
+[\Delta_{13}(d-\Delta_{13})+m^2]f
= t^{\Delta_3}
\ee
The particular solution that corresponds to the actual value
of the integral  (\ref{scalarintegral}) is selected by the following asymptotic
conditions on $f(t)$:
\begin{enumerate}
\item Since $I(w)$ is perfectly regular at $\vec w= 0$,
$f(t)$ must be smooth as $t \to 1$.
 \item In the limit
$w_0 \to 0$ we have from (\ref{scalarprophyper})
and (\ref{u}) that $I(w) \sim w_0^\Delta$, which implies
$f(t) \sim t^{\frac{\Delta-\Delta_{13}}{2}}$ as $t \to 0$. (Recall
that we are considering the case $\Delta= \Delta_{+}$).
\end{enumerate}

The differential operator in (\ref{scalardiffequ})
is closely related to the hypergeometric
operator,
and we will discuss this shortly, but for the cases of interest
we can find a particular solution of
the equation 
more quickly if we convert it to a recursion relation.
To do this we assume the series representation 
\be \label{scalarseries}
f(t)= \sum_k a_k t^k \,. 
\ee
Upon
substitution in (\ref{scalardiffequ}) 
we find a recursion relation for the coefficients
which works downwards in $k$. We can consistently
set $a_{k} =0$ for $k \geq \Delta_3$. We then get      
\bea
a_k &= &0     \qquad         {\rm for}\; k \geq \Delta_3 \label{scalarsol1}
\\
a_{\Delta_3-1}& =&  \frac{1}{4(\Delta_1 -1)(\Delta_3 -1)}
 \\
a_{k-1} &=& \frac{(k-\frac{\Delta}{2}+\frac{ \Delta_{13} }{2} )(k-\frac{d}{2}
+\frac{\Delta}{2}+\frac{\Delta_{13}}{2})}{(k-1)(k-1+\Delta_{13})} \; a_k
\label{scalarsol3}
\eea
Note that $k$ need not take integer values, rather $k=\Delta_3 + l$
with $l$ integer but $\Delta_3$ arbitrary.
We now observe that the series terminates at the positive value\footnote{ One may also consider solutions which terminate because the second
factor in the numerator of (\ref{scalarsol3}) vanishes, which gives a lower
value of $k_{{\rm min}}$. We have not studied this possibility 
since it does not
satisfy the required behavior as $w_0 \to 0$.}
$k_{{\rm min}}=(\Delta-\Delta_{13})/2 \leq  k_{{\rm max}} = \Delta_3-1$ provided that 
$\Delta_1 +\Delta_3 -\Delta$ is a positive even integer. 
If 
(and only if) 
this condition 
is satisfied, (\ref{scalarseries}--\ref{scalarsol3}) give a
well--defined
particular solution of (\ref{scalardiffequ})
with the required
asymptotic properties. We will shortly prove its uniqueness.

It is pleasant to observe
that the condition for terminating series is satisfied
for all the cases that occur in type 
IIB $AdS_5 \times S_5$ supergravity \cite{vannieuv}
due
to restrictions on trilinear couplings from $SU(4)$ symmetry
\cite{slanski, dhfscalar}\footnote{$SU(4)$ group theory also allows
the case $\Delta_1+\Delta_3-\Delta=0$, for which
our particular solution is either ill--defined or non--terminating.
(In this latter case it is singular at $t=1$.)
However it appears that in these cases the trilinear
supergravity coupling contains derivatives, and the relevant
integral can be transformed to integrals
 obeying the termination condition,
see the Appendix of 
\cite{dhfscalar}.}. 
In this paper we will only consider the terminating case.

We can easily prove uniqueness of the solution
(\ref{scalarseries}--\ref{scalarsol3}) by
showing that any combination of the two homogeneous
solutions of (\ref{scalardiffequ}) fails to satisfy
the asymptotic requirements on $f(t)$. By making
the change of variable $x=1/t$, we can write the homogeneous
equation as
\be
x(1-x) f''(x) +[1-\Delta_{13}-(1-\Delta_{13}+\frac{d}{2})x]f'(x)
-\frac{1}{4}(\Delta_{13}-\Delta)(\Delta_{13}+\Delta-d)f(x)=0
\ee
which is the hypergeometric equation of
parameters $a=\frac{\Delta-\Delta_{13}}{2}$, 
$b=\frac{d-\Delta-\Delta_{13}}{2}$, $c=1-\Delta_{13}$. 
Two independent homogeneous solutions
of (\ref{scalardiffequ})
are then given by \cite{bateman}
\bea
f_1(t)&=& t^{\frac{\Delta-\Delta_{13}}{2}} F(\frac{\Delta-\Delta_{13}}{2},
\frac{\Delta+\Delta_{13}}{2}; \Delta-\frac{d}{2} +1;t)
\\
f_2(t)&=&F(\frac{\Delta-\Delta_{13}}{2}, \frac{d-\Delta-\Delta_{13}}{2};
\frac{d}{2};1-\frac{1}{t})
\eea
It is easy to see that $f_1$ is singular for $t \to 1$, while
$f_2$ is regular in the same limit. We must then reject
$f_1$ based on the first asymptotic condition stated above.
For $t \to 0$, $f_2(t) \sim t^{\frac{d-\Delta-\Delta_{13}}{2}}$, 
which violates the second asymptotic condition. $f_2$ 
scales at small $t$ with the rate 
corresponding to the ``irregular''
choice of boundary condition for the bulk scalar, {\it i.e.} 
$\Delta=\Delta_{-}$.
The value of the integral (\ref{scalarintegral})
for $\Delta=\Delta_{-}$ could be obtained in the terminating case
by adding to  the particular solution
(\ref{scalarseries}--\ref{scalarsol3}) a multiple of  $f_2$.

We now make contact with the results of \cite{dhfscalar}.
The restriction of $\Delta_1+ \Delta_3-\Delta$ to positive even integers 
agrees
with the condition stated after (3.22) of \cite{dhfscalar}
 for termination of the 
(transformed) hypergeometric series in  (3.10) or (3.11).
The integral in (3.11) then yields
a polynomial expression wich precisely
agrees with (\ref{scalarseries}--\ref{scalarsol3}).
 Note that the
integral $I(w)$ was called $R(w)$ in  \cite{dhfscalar}. 

We can finally assemble the result for the initial amplitude
(\ref{scalarw}). From (\ref{scalarform}), (\ref{If}--\ref{t}),
(\ref{scalarseries}),
 inverting back
to the original coordinates $\vec x_i$ (see (\ref{inversion})), we have
\bea
&&S(\vec x_1, \vec x_2,\vec x_3,\vec x_4)= \\
&&
\sum_{k_{\rm min}}^{k_{\rm max}}
a_k \, |\vec x_{13}|^{-2\Delta_3+2k}
\int { d^{d+1}w \over w^{d+1}_0}\, 
K_{\Delta_1-\Delta_3+k}(w,\vec x_1) \,K_{k}(w,\vec x_3)
\,K_{\Delta_2}(w,\vec x_2)\,K_{\Delta_4}(w,\vec x_4) \nonumber\,,
\eea
{\it i.e.} the exchange amplitude reduces to a finite sum
of scalar quartic graphs. The analytic properties
of these quartic graphs have been extensively
studied \cite{canadians, august, dhfgauge,dhfscalar,march99}.
In particular 
asymptotic expansions in terms of conformally invariant variables
are available. We  refer the reader to Section 6 and to Appendix A
of \cite{march99} for a self--contained derivation of
these expansions and of many other useful identities.

\section{Vector exchange}
\setcounter{equation}{0}

The basic procedure for vector and tensor exchange integrals is the same
as in the scalar case. We use the wave equation satisfied by the 
bulk--to--bulk propagator to turn the integral into  an
inhomogeneous second 
order differential
equation for scalar functions of $t= (w_0)^2/w^2$ and then obtain the
             particular solution with required asymptotics by a
recursion relation. 
The choice
of a suitable ansatz which expresses the vector or tensor valued integral in
terms of scalar functions and the action of the wave operator on that ansatz
are more complicated than in the scalar case.

For vector exchange we study the integrals
\be \label{vectorw}
V(\vec x_1, \vec x_2, \vec x_3, \vec x_4)=
\int  \frac{d^{d+1}w}{w_0^{d+1}}\,
A_{\mu}(w,\vec x_1 ,\vec x_3) \;g^{\mu \nu}(w)\;K_{\Delta_2}(w, \vec x_2) \frac{ \stackrel{\leftrightarrow}{\partial} 
}{\partial w_{\mu}}  
K_{\Delta_2}(w, \vec x_4) 
\ee
where
\be \label{vectorintegral}
A_{\mu}(w,\vec x_1 ,\vec x_3)=
\int \frac{d^{d+1}z}{z_0^{d+1}}\,
G_{\mu \nu'}(w,z) \,
g^{\nu' \rho'}(z)
\, K_{\Delta_1}(z, \vec x_1) \frac{ \stackrel{\leftrightarrow}{\partial} 
}{\partial z_{\rho'}}  
K_{\Delta_1}(z, \vec x_3) 
\ee
Note that we use unprimed indices for the $w$ coordinate and primed 
indices for $z$.
The only information we need about the bulk--to--bulk propagator is the
defining wave equation, namely
\be \label{vectorpropequ}
-\frac{1}{\sqrt{g}} \partial_{\mu}(\sqrt{g} g^{\mu \lambda}
\partial_{[\lambda}G_{\rho] \nu'}(w,z) ) + m^2 G_{\rho \nu'}(w,z)=
g_{\rho \nu'} \delta(w,z) + \partial_{\mu} \partial_{\nu'}\Lambda(u)
\ee
where the first term is the Maxwell operator and the second is the mass term.
The pure gauge term appears on the right side only for $m^2=0$ because the
operator is then non--invertible. For $m^2 \neq 0$,
this is the appropriate wave equation for the massive vector fields of 
type IIB supergravity on $AdS_5\times S_5$  \cite{vannieuv}. 
We have also assumed 
that vectors couple to the conserved current 
formed from the two bulk--to--boundary
propagators in (\ref{vectorintegral}). This is certainly 
the case for massless gauge bosons,
and we restrict attention to conserved current sources for  massive KK vectors  also. The method can be extended to include more general sources.

The propagator transforms as a bitensor under inversion, so the integral 
transforms to the inverted frame as \cite{april}
\be
A_{\mu}(w,\vec x_1 ,\vec x_3)= |\vec x_{13}|^{-2\Delta_1}
\frac{1}{w^2} J_{\mu \nu}(w) I_{\nu}(w'-\vec x'_{13})
\ee
where $ J_{\mu \nu}(w)=\delta_{\mu \nu} -2 \, w_\mu w_\nu /w^2 $ and 
\be
I_\mu(w)=\int \frac{d^{d+1}z}{z_0^{d+1}}\,
G_{\mu}^{\;\; \nu'}(w,z) \,
\, z_0^{\Delta_1} \frac{ \stackrel{\leftrightarrow}{\partial} }{\partial z_{\nu'}}  
\left( \frac{z_0}{z^2}\right)^{\Delta_1}  \,.
\ee

We now need a suitable ansatz for the vector function $I_\mu(w)$. Scale
symmetry and $d$--dimensional Poincar{\'e} symmetry suggest the form
\be \label{vectoransatz}
I_\mu(w)= {w_\mu \over w^2} f(t) +{\delta_{\mu0} \over w_0} h(t)
\ee
However, the second term can be dropped because of the following argument. The
first step is the observation that $D^\mu I_\mu(w)= 0$. This follows because
the divergence $D^\mu G_{\mu \nu'}(w,z)$ is a rank 1 bitensor in a maximally
symmetric space, and must then be proportional to the
only independent rank 1 bitensor \cite{allenjacobson}, 
namely $\partial_{\nu'} u$, times a scalar function of $u$. 
$D^\mu G_{\mu \nu'}(w,z)$ can 
then be expressed as a $z$--derivative of a scalar function:
\be
D^\mu G_{\mu \nu'}(w,z)=\partial_{\nu'}u\, g(u)=
\partial_{\nu'}\left( \int g(u) \right) \,.
\ee
This 
gradient term\footnote{For massive vectors, the field equation
implies that the gradient term is proportional to
$\partial_{\nu'} \delta(w,z)$.}
 can then be partially integrated in the $z$--integral for 
$D^\mu I_\mu(w)= 0$ and vanishes by current conservation. The
divergence can now be applied to the ansatz (\ref{vectoransatz}), 
which gives
\be
0=D^\mu I_\mu(w)= D^\mu\left( {w_\mu \over w^2} f(t)\right)
 + D^\mu\left( {\delta_{\mu0} \over w_0} h(t) \right)
\ee
The first term vanishes identically, while the second term leads to
a separable first order homogeneous equation for $h(t)$. The non--trivial
solution is singular as $t \rightarrow 1$ and must be rejected, since
we see by inspection of (\ref{vectorintegral}) 
that $I_\mu(w)$ is regular there. Thus we
have proven that $h(t)=0$ and we can use the representation
\be
I_\mu(w)= {w_\mu \over w^2} \,f(t).
\ee

We now apply the wave operator to $I_\mu(w)$, and use (\ref{vectorpropequ})
 under the integral
sign (the gauge term vanishes when integrated by parts). The result is the
equation 
\be
-\frac{1}{\sqrt{g}} \partial_{\mu}\left( \sqrt{g} g^{\mu \lambda}
\partial_{[\lambda} \left( \frac{w_{\rho]}}{w^2} f(t)
\right) \right) + m^2 \, \frac{w_{\rho}}{w^2}\,f(t)
=-2 \Delta_1 {w_\rho \over w^2} \,t^{\Delta_1} \,.
\ee
It is now straightforward, although complicated, to calculate the result of
the action of the Maxwell operator on the left side, and this leads to the
differential equation
\be \label{vectordiffequ}
4t^2(t-1)f'' +4t(2t + {{d-4}\over 2}) f' + m^2 f = -2\Delta_1 
\,t^{\Delta_1} \,.
\ee
This inhomogeneous differential equation is clearly
of the same type as (\ref{scalardiffequ})
for scalar exchange.
We thus proceed in the same way  by looking
 for a particular solution of the form
\be \label{vectorseries}
f(t)=\sum_k a_k t^k \, .
\ee
with $k \in \{k_{{\rm min}}, k_{{\rm min}}+1, \dots , k_{{\rm max}} \}$.
We find:
\bea
a_k &= &0     \qquad         {\rm for}\; k \geq \Delta_1 \label{vectorsol1}
\\
a_{\Delta_1-1}& =&  \frac{1}{2(\Delta_1 -1)}
 \\
a_{k-1} &=& \frac{2k(2k+2-d)-m^2}{4(k-1)k} \; a_k
\label{vectorsol3} \,.
\eea
The series terminates at $0< k_{{\rm min}}=\frac{d-2}{4}+\frac{1}{4}
\sqrt{(d-2)^2+4m^2} \leq  k_{{\rm max}} =\Delta_1-1$
provided that $ k_{{\rm max}}-k_{{\rm min}}$ is integer and $\geq 0$.
 It is easy to check, in analogy with the scalar case,
that if this condition is obeyed, (\ref{vectorseries}--\ref{vectorsol3})
define the unique particular solution of (\ref{vectordiffequ}) 
with the correct asymptotic
properties to correspond to the actual value of the integral 
(\ref{vectorintegral}). 
For $m^2 =0$ and $d=2$, the coefficient $a_{k_{\rm min} }=a_{0}$
is infinite, a signal that this case requires special attention.

We now consider the application of these results to
the $AdS_5 \times S_5$ compactification of IIB supergravity.
From Table III in \cite{vannieuv}, we see that the allowed
values of the mass for KK vectors are $m^2=(l-1)(l+1)$, with $l$
integer $\geq 1$. The termination condition
then requires $\Delta_1-1-( 1/2 +l/2)$ be a non--negative integer,
which  restricts $l$ to be {\it odd} and $< 2\Delta_1-1$.
It can be shown that $SU(4)$ selection rules \cite{slanski}
enforce $l$ odd, $l \leq 2\Delta_1-1$. In fact, the value of $l$
correlates to the quadrality of the $SU(4)$ representation
of the vector field,
the quadrality is 0 or 2 for $l$ odd or even. Since
scalar fields come in representations 
with quadrality
2 or 0, and we are assuming two {\it equal} scalar 
fields $\Delta_1=\Delta_3$, imposing
that the sum of the quadralities in the trilinear coupling
is 0 mod 4 forces $l$ to be odd. The inequality
$l \leq 2 \Delta -1$ is the standard ``Clebsch--Gordon''
triangle inequality.
We thus observe the nice phenomenon that precisely 
for the cases allowed in the supergravity
we get  terminating series  for the vector exchange $z$--integrals\footnote{
One possible exception to this is the marginal case
$l = 2 \Delta_1-1$. We would expect, in analogy to the scalar
exchange, that this case
occurs in the actual supergravity theory with a different coupling. 
It would be nice to check this explicitly
from the supergravity lagrangian.}.

We now wish to compare with the results of \cite{dhfgauge},
where the massless vector exchange was computed. For $m=0$,
the termination condition requires $\Delta_1 -d/2$ be a non--negative
integer, which is in particular satisfied for 
$d$ even and  $\Delta_1$ integer
satisfying the unitarity bound. This is the condition stated
in \cite{dhfgauge} after (3.21) for the 
$z$--integral  (3.20) to reduce to a finite sum of elementary
terms. Comparison with the results of the present
paper shows perfect agreement.

\section{Graviton exchange}
\setcounter{equation}{0}

The tensor exchange integral is more complicated than 
previous cases, although the new method is still considerably simpler
than that of previous work \cite{march99}.
We start with the
integral
\be \label{gravitonw}
G(\vec x_1, \vec x_2, \vec x_3, \vec x_4)=
\int \frac{d^{d+1}w}{w_0^{d+1}}\,
A^{\mu\nu}(w,\vec x_1 ,\vec x_3)\;T_{\mu \nu}(w,\vec x_2, \vec x_4)
\ee
where
\be \label{gravitonz}
A_{\mu\nu}(w,\vec x_1 ,\vec x_3)=\int \frac{d^{d+1}z}{z_0^{d+1}}\,
G_{\mu \nu \mu' \nu'}(w,z) T^{\mu' \nu'}(z, \vec x_1, \vec x_3) \,.
\ee
The stress tensor governing the couplings of the bulk graviton to
scalar fields of equal dimensions $\Delta_1=\Delta_3
=\frac{d}{2}+\frac{1}{2}\sqrt{{d^2}+4 m_1^2}$ is given by
\bea \label{stress}
T^{\mu'\nu'}(z, \vec x_1, \vec x_3) &= & ~ D^{\mu'} K_{\Delta_1} (z, \vec x_1)
 D^{\nu'} K_{\Delta_1} (z,\vec  x_3) \\
&& - \frac{1}{2}\,g^{\mu' \nu'} \bigl [ D_{\rho'} K_{\Delta_1} (z,\vec x_1) 
D^{\rho'} K_{\Delta_1} 
(z,\vec x_3)
+ m_1^2 K_{\Delta_1} (z,\vec x_1) K_{\Delta_1} (z, \vec x_3) \bigr ]\, . 
 \nonumber
\eea
The graviton propagator $G_{\mu \nu \mu' \nu'}(w,z)$ was discussed extensively in \cite{january}, but 
the main property needed here is the (Ricci form) of its wave equation,
namely
\bea \label{gravitonpropequ}
W_{\mu \nu}^{ \;\;\;\lambda \rho} G_{\lambda \rho \mu ' \nu '} &\equiv 
 &-  D^\sigma D_\sigma G_{\mu \nu \mu ' \nu '}
  -  D_\mu D_\nu G_{\sigma} {}^\sigma {}_{\mu ' \nu '}
  +  D_\mu D^\sigma G_{ \sigma \nu  \mu ' \nu '} \\
  &&+  D_\nu D^\sigma G_{\mu \sigma  \mu ' \nu '}
-2(G_{\mu \nu \mu ' \nu '} - g_{\mu \nu} G_\sigma {}^\sigma
{}_{\mu ' \nu'}) \nonumber\\  
 &=&
 \Bigl(g_{\mu \mu'}g_{\nu \nu'} +g_{\mu \nu'}g_{\nu \mu'} -  {2\over d-1}
g_{\mu\nu}g_{\mu' \nu'}\Bigr)
\delta(z,w)  + D_{\mu '} \Lambda _{\mu \nu \nu'} + D_{\nu '}\Lambda _{\mu \nu 
\mu'}
\nonumber
\eea
The form of the pure diffeomorphism $\Lambda_{\mu \nu \nu'}$
need not be discussed (see 
\cite{january}) since it drops out when the wave operator is
applied to the integral using covariant conservation of $T_{\mu'\nu'}$.
The transformation to inverted coordinates gives
\be
A_{\mu \nu}(w, \vec x_1, \vec x_3) = |\vec x_{13}|^{-2 \Delta_1}
\frac{1}{(w^2)^2}\,J_{\mu \lambda}(w)
J_{\nu \rho}(w)\,I_{\lambda \rho}(w'-\vec x'_{13})
\ee
with the tensor integral 
\bea \label{gravitonintegral}
I_{\mu \nu}(w) &=& \int \frac{d^{d+1}z}{z_0^{d+1}}\,
G_{\mu \nu \mu' \nu'}(w,z) \left[
 ~ D^{\mu'} z_0^{\Delta_1} D^{\nu'} \left( \frac{z_0}{z^2} \right)^{\Delta_1}
 \right. \\
&& \left. - \frac{1}{2}\,g^{\mu' \nu'} \bigl [ D_{\rho'} z_0^{\Delta_1}  
 D^{\rho'}  \left( \frac{z_0}{z^2} \right)^{\Delta_1}
+ m_1^2  z_0^{\Delta_1}   \left( \frac{z_0}{z^2} \right)^{\Delta_1}
\right] \, . 
 \nonumber
\eea
which we shall now study.
The first step is to find a suitable ansatz for this integral with 
independent tensors multiplying scalar functions of $t=(w_0)^2/w^2$. The 
most suitable basis appears to be
\be \label{gravitonansatz}
I_{\mu \nu}(w)= g_{\mu \nu} \,h(t) + \frac{\delta_{0\mu}\delta_{0\nu} }{w_0^2}
\, \phi(t)
+D_\mu D_\nu X(t) + D_{\{ \mu } \left( \frac{\delta_{ \nu \}0 } }{w_0}\,
Y(t) \right)
\ee
where $\{\, \}$ denotes symmetrization. The last two terms 
in (\ref{gravitonansatz}) are pure diffeomorphisms
and 
depend on the gauge
choice for the graviton propagator. They are annihilated
by the Ricci wave operator and are thus not determined by the present
technique. On the other hand they have no physical effect, since they
drop out of the final $d^{d+1}w$ integral which contains another conserved 
stress tensor. 

We now apply the Ricci wave operator to $I_{\mu\nu}$ in 
(\ref{gravitonintegral})
  and use  (\ref{gravitonpropequ}) to
obtain, after some simplification,
\be \label{Ricciansatz}
W_{\mu \nu}^{ \;\;\;\lambda \rho} \left[
 g_{\lambda \rho} h(t) + \frac{\delta_{0\lambda}\delta_{0\rho}}{w_0^2}\phi(t)
\right] = 2 \tilde T_{\mu \nu}
\ee
with
\bea \label{Ttilde}
 2 \, \tilde T_{\mu \nu} &=& \partial_\mu  w_0^{\Delta_1} \,\partial_\nu
\left( \frac{w_0}{w^2} \right)^{\Delta_1}
+ \frac{2}{d-1}\,m_1^2 \;g_{\mu \nu}\,t^{\Delta_1} +(\mu \leftrightarrow \nu)
\\
&= &2 \Delta_1^2\,\left( \frac{\delta_{\mu 0} \delta_{\nu 0}}{w^2}
-\frac{w_0(\delta_{\mu 0} w_\nu+ \delta_{\nu 0} w_\mu)}{(w^2)^2}
\right)t^{\Delta_1-1} +\frac{2 m_1^2}{d-1} \, g_{\mu \nu}\, t^{\Delta_1}
\nonumber
\eea
The major task is to apply the wave operator to the two tensors on the left
side. The courage and fortitude necessary for this task are stimulated by 
the previous
successes of the method in Sections 2 and 3.
The task is eased to some extent
by defining the ``vector'' 
\be
P_\mu \equiv {\delta_{\mu 0} \over w_0}
\ee
which satisfies
\be
P_{\mu}P^{\mu}=1\,, \qquad D_\mu P_\nu = -g_{\mu \nu} + P_\mu P_\nu\,,
\qquad D^\sigma P_\sigma = -d\,.
\ee
We simply give the results of these calculations:
\bea
W_{\mu \nu}^{ \;\;\;\lambda \rho} \left[
 g_{\lambda \rho} h(t) \right] &=&g_{\mu \nu}
\, \left[ 4t^2(t-1) h''(t)+4t \left(t-1+d/2 \right)h'(t) +2d \,h(t)
 \right] \\
&& + (-d+1)D_\mu D_\nu h(t) \nonumber 
\eea
\bea
W_{\mu \nu}^{ \;\;\;\lambda \rho}
\left[ \frac{\delta_{0\lambda}\delta_{0\rho}}{w_0^2}\, \phi(t) \right]
&=& g_{\mu \nu} 
\left[ 4t(t-1)\,\phi'(t)+2d\,\phi(t)              \right] + \\
&& \frac{\delta_{0\{\mu} w_{\nu\}}\,w_0}{(w^2)^2} 
\left[ 4t(t-1) \phi''(t)+(8t+2d-8)\phi'(t)         
\right]+
\nonumber \\
&& \frac{\delta_{0 \mu} \delta_{0 \nu}}{w^2} 
\left[ 4t(1-t) \phi''(t)+(-8t-2d +8)\phi'(t)  \right] 
\nonumber \\
&& - D_\mu D_\nu \phi(t) \,.               \nonumber 
\eea

The remaining task is to use the information in the four independent tensor
contributions to (\ref{Ricciansatz}). We have an overdetermined
system of 4 equations for
2 unknown functions, so compatibility of the system will provide 
a check of the method.

The tensor $w_\mu w_\nu$ does not appear on the right
side, and it appears on the left hand side
only in the expansion of $D_\mu D_\nu h$ and  $D_\mu D_\nu \phi$
\be
D_\mu D_\nu A(t)
=\frac{w_\mu w_\nu}{(w^2)^2}\,\left( 4t^2 A''(t)
+8t A'(t) \right)+\dots
\ee
with $A(t)=-\phi(t) +(1-d) h(t)$.
So we get the condition
\be \label{condition}
 h(t) =\frac{1}{1-d} \,\phi(t) \, ,
\ee
where we have 
chosen the trivial homogeneous solution of  $4t^2A''+8tA'=0$
because any other solution would be incompatible with 
the asymptotic behavior of the integral (\ref{gravitonintegral}),
which vanishes as $t \to 0$.

Equating the contributions of the tensor structure 
$\delta_{0 \mu} \delta_{0 \nu}/{w^2}$  to the l.h.s. and r.h.s.
of  (\ref{Ricciansatz}) we get:
\be \label{deltaequ}
 4t(1-t) \phi''(t)+(-8t-2d +8)\phi'(t)=2\Delta_1^2 \,t^{\Delta_1-1}\,.
\ee
The equation obtained from the tensor 
 $\delta_{0\{\mu} w_{\nu\}}\,w_0/(w^2)^2$  differs
from  (\ref{deltaequ}) just in overall sign, so
the first of the two required compatibility conditions is satisfied.

Finally collection of the terms proportional to $g_{\mu \nu}$ gives
\be \label{hequ}
4t^2(t-1) h''(t)+4t \left(t-1+d/2 \right)h'(t) +2d \,h(t)
+ 4t(t-1)\,\phi'(t)+2d\,\phi(t)  =\frac{2m_1^2}{d-1} \, t^{\Delta_1}
\ee
Compatibility of this last equation with
(\ref{condition}) and (\ref{deltaequ}) is readily
shown as follows. 
Let us 
eliminate $h(t)$ from (\ref{hequ}) using (\ref{condition}), 
and multiply the resulting equation by $(d-1)/t$. We obtain
\be \label{resulting}
 4t(1-t) \phi''(t)+\left[(-8+4d)t-6d +8\right]\phi'(t)+\frac{2d(d-2)}{t}
\phi(t)
=2 m_1^2 \,t^{\Delta_1-1}\,.
\ee
Subtracting (\ref{resulting}) from  (\ref{deltaequ}),
and using $m_1^2 =\Delta_1(\Delta_1-d)$, 
we get a first order equation for $\phi$
\be\label{first}
4t(1-t)\phi'(t)-2(d-2)\phi(t)=2\Delta_1 t^{\Delta_1}\,,
\ee 
which is obviously compatible with  (\ref{deltaequ}), the latter
just being the derivative of the first order equation (\ref{first}).
We thus conclude that the system of 4 differential
equations is consistent and all of its information is contained
in the two simple equations (\ref{first}) and (\ref{condition}).

To find the particular solution of (\ref{first}), as in the
scalar and vector cases we consider an ansatz of the form
\be \label{gravitonseries}
\phi(t) = \sum_k a_k t^k
\ee
with a finite span of values of $k$,
$k \in \{ k_{{\rm min}},  k_{{\rm min}}+1,\dots,  k_{{\rm max}} \}$. We find:
\bea
a_k &= &0     \qquad         {\rm for}\; k \geq \Delta_1 \label{gravitonsol1}
\\
a_{\Delta_1-1}& =&  -\frac{\Delta_1}{2(\Delta_1 -1)}
 \\
a_{k-1} &=& \frac{k+1 -\frac{d}{2}}{k-1} \; a_k
\label{gravitonsol3} \,.
\eea
The series terminates at $k_{{\rm min}}=d/2-1 \leq k_{{\rm max}}=
\Delta_1-1$ provided $\Delta_1-d/2$ is a non--negative integer and $d>2$.

Actually it is quite trivial to integrate
(\ref{first}), and it is instructive to compare the direct
solution with the solution by recursion. The general solution of
(\ref{first}) is
\be
\phi(t)= -\frac{\Delta_1}{2} \,\left( \frac{t}{t-1}
 \right)^{\frac{d}{2}-1} \,\left\{ \int_1^t dt' t'^{\Delta_1-\frac{d}{2}}
(t'-1)^{\frac{d}{2}-2} +c \right\}
\ee
where $c$ is arbitrary. Assume, for simplicity, that $d$ is an even
integer\footnote{If $\frac{d}{2}-2 = \alpha$ is not an integer, the same
conclusions follow if one makes the successive changes of
variable $u=t'-1$ and then $u=v^\beta$ with $\beta=1/\alpha$.}.
For $d > 2$ one must choose $c=0$ to avoid a singularity at
$t=1$. The integral solution is then a polynomial in $t$ if and only
if $\Delta_1-d/2$ is a non--negative integer. For $d=4$ and
$\Delta$ integer, the result is the simple polynomial
\be
\phi(t) = -\frac{\Delta_1}{2(\Delta_1-1)} \,(t+t^2 + \dots+t^{\Delta_1-1})\,.
\ee
For $d=2$, there is an unavoidable ${\rm ln}(t-1)$. 
This is another indication that the case $d=2$ requires special
consideration.

The acid test of the new method is to compare with previous
results which were given in \cite{march99}. 
The most direct comparison available is for $d=4$ and
$\Delta_1 =4$ for which results were given in
(5.64) and (5.65) of \cite{march99}. Agreement is perfect
after different normalizations are taken into account. For 
general $\Delta_1$ and $d$ the new method gives a much more
coincise result for the amplitudes.

\section{Higher point functions}
\setcounter{equation}{0}

The methods developed in the preceding sections for the calculation of the 
$z$-integrals involving two bulk-to-boundary propagators may be generalized 
to the case where the bulk-to-bulk propagator is integrated with an 
arbitrary
number $n$ of bulk-to-boundary propagators. This generalization will be 
required
when the effects of supergravity couplings of the form $\phi^{n+1}$ are taken 
into 
account. This will indeed be the case when AdS/CFT amplitudes are evaluated
to higher order in the supergravity coupling $\kappa \sim 1/N$. 

For simplicity, we shall restrict attention here to the case of scalar
bulk-to-boundary and scalar bulk-to-bulk propagators only.  We shall assume
the dimension $d$ of AdS space and of the scaling dimensions $\Delta_i$ of 
all fields to be integers, subject to the unitarity bound $\Delta _i \geq d/2$.
Furthermore, we shall assume that at any given interaction vertex, the 
dimensions of the fields satisfy the standard triangle inequality, which,
for $AdS_5 \times S^5$ results directly from the $SO(6)$ R-symmetry.

The starting point is the $z$-integral, defined by
\be
R(w)= \int dz \sqrt g ~ G_\Delta (u) \prod _{i=1} ^n 
\biggl (\frac{z_0}{z_0^2 + (\vec z - \vec x_i )^2} \biggr ) ^{\Delta _i}
\ee
where $G_\Delta(u)$ is the scalar propagator of dimension $\Delta$ and mass
$m^2 = \Delta (\Delta -d)$, obeying (\ref{scalardelta}), and $u$ is a function 
of $z$ and $w$. From (\ref{scalardelta}), it is clear that $R(w)$ satisfies the
following differential equation
\be \label{reqn}
(\Box - m^2) R(w)=  \prod _{i=1} ^n 
\biggl (\frac{w_0}{w_0^2 + (\vec w - \vec x_i )^2} \biggr ) ^{\Delta _i}
\, .
\ee
The source term on the r.h.s. may be re-expressed as an integral over Feynman 
parameters $\alpha _i$, $i=1,\cdots ,n$ of a rational function with a single 
denominator,
\be 
 \prod _{i=1} ^n 
\biggl (\frac{w_0}{w_0^2 + (\vec w - \vec x_i )^2} \biggr ) ^{\Delta _i}
=
\frac{\Gamma (\delta)}{\prod _i \Gamma (\Delta _i)}
\prod _{i=1}  ^n \int _0 ^1 d\alpha _i ~ \alpha _i ^{\Delta _i-1}
\frac{\delta (1 - \sum _{i=1} ^n \alpha _i) ~ w_0^\delta}{(w_0^2 + (\vec w - 
\vec v)^2 + \mu ^2)^\delta}
\, .
\ee
Here, we have defined the abbreviations 
\bea
\delta & = & \Delta _1 + \Delta _2 + \cdots +\Delta _n \\
\vec v & = & \alpha _1 \vec x_1 + \alpha _2 \vec x_2 + \cdots 
+ \alpha _n \vec x_n \\
\mu^2 & = & - \vec v^2 + \alpha _1 \vec x_1 ^2 + \alpha _2 \vec x_2^2 +
\cdots + \alpha _n \vec x_n ^2 
\eea
Here, it is understood that both $\vec v$ and $\mu^2$ are functions of the 
Feynman parameters $\alpha _i$. Using the linearity of  (\ref{reqn}), the 
solution for $R(w)$ may be obtained as follows
\be
R(w) = \frac{\Gamma (\delta)}{\prod _i \Gamma (\Delta _i)}
\prod _{i=1}  ^n \int _0 ^1 d\alpha _i ~ \alpha _i ^{\Delta _i-1}
\delta (1 - \sum _{i=1} ^n \alpha _i) S(w-\vec v;\delta; \mu)
\, ,
\ee
where the scalar function $S(w;\delta;\mu)$ satisfies the differential equation
\be \label{key}
(\Box - m^2) S(w;\delta;\mu) = \frac{w_0^\delta}{(w^2 + \mu ^2)^\delta}
\, .
\ee
The key problem is thus to solve for (\ref{key}) as a function of $w$. Once the 
function $S$ is known, the function $R(w)$ can be found by carrying out the 
remaining Feynman integrals. As we shall see, under certain restrictions on the
dimensions $\Delta$, $\Delta _i$ and $d$, the function $S$ will be polynomial
in $w_0 /(w^2 +\mu^2)$, and thus the Feynman integrals to be calculated are
of a standard type.

To solve for (\ref{key}), we begin by noticing that the operator $\Box $ 
applied to a power of $w_0 /(w^2 +\mu^2)$ yields a function of the same type.
Actually, one may easily show a slightly more general formula that may be
useful to treat the cases of vector and tensor bulk-to-bulk propagators, 
\be
\Box \frac{w_0^\ell}{(w^2 +\mu^2)^k}
=
\ell (\ell -d) \frac{w_0^\ell}{(w^2 +\mu^2)^k}
+4k (k -\ell) \frac{w_0^{\ell+2}}{(w^2 +\mu^2)^{k+1}}
-4k(k+1) \mu ^2 \frac{w_0^{\ell+2}}{(w^2 +\mu^2)^{k+2}}
\, .
\ee
Remarkably, for the case at hand, where $k=\ell$, this double recursion 
simplifies.
Subtracting also the mass term $m^2=\Delta (\Delta -d)$, as 
will be needed for the resolution of equation (\ref{key}),  we find the simple 
recursion relation 
\be \label{recursion}
(\Box  - m^2) \frac{w_0^\ell}{(w^2 +\mu^2)^\ell}
=
(\ell - \Delta) (\ell +\Delta -d) \frac{w_0^\ell}{(w^2 +\mu^2)^\ell}
-4\ell (\ell +1) \mu ^2 \frac{w_0^{\ell+2}}{(w^2 +\mu^2)^{\ell+2}}
\, .
\ee
It remains to solve (\ref{key}).

We now follow the spirit of previous sections and investigate 
solutions of (\ref{key}) which can be expressed as a finite series of powers
of the variable $w_0/(w^2+\mu^2)$. Using (\ref{recursion}) one sees that this
is possible if the highest power is $l_{\rm max} =\delta -2$ with lower powers
given by $l=l_{\rm max} -2j$ 
where $j$ is a positive integer. The series terminates
at $l_{\rm min} =\Delta$ provided that $\delta - \Delta -2= 2\ell_0$ is a
non--negative even integer\footnote{There is another possible solution which
 terminates at $l_{{\rm min}}=d-\Delta$.
We do not study this since it is superceded by the previous solution if
$\Delta$ is an integer, as is the case for scalar fields in Type IIB 
supergravity.}.   (For $n=2$ this condition coincides with the
condition for a terminating
 solution in Section 2). Substituting (\ref{recursion})
in (\ref{key}) one finds that the solution takes the form
\be
S(w;\delta;\mu) = \sum _{\ell =0} ^{\ell_0} C_\ell (\mu) 
\frac{w_0^{\Delta + 2\ell}}{(w^2+\mu^2)^{\Delta + 2\ell}}
\ee
with the recursion relation for the coefficients,
\bea
C_{\ell -1} (\mu) & = & \frac{1}{\mu^2} 
\frac{2\ell (2 \ell + 2 \Delta -d)}{4(\Delta +2\ell-2)(\Delta +2\ell -1)} 
C_\ell (\mu) \\
C_{\ell_0} & = & - \frac{1}{\mu^2} \frac{1}{4(\delta -1)(\delta-2)} 
\eea
This recursion relation is easily solved and one finds
\be
C_\ell (\mu) = - \frac{1}{4} \mu ^{2\ell +\Delta -\delta}
\frac{\Gamma(\half(\delta -\Delta)) \Gamma (\half(\delta -\Delta -d))
\Gamma (\Delta + 2\ell)}{\Gamma(\delta) \Gamma(\ell+1) \Gamma(\ell +\Delta 
+1-d/2)}
\ee
Remarkably, the conditions for polynomial solutions are precisely obeyed
thanks to the R--symmetry selection rules of $AdS_5\times S^5$ supergravity.

\section*{Acknowledgments}

It is a pleasure to acknowledge useful conversations with
Samir Mathur and Alec Matusis.

The research of E.D'H is supported in part by NSF Grant
PHY-95-31023, 
D.Z.F.  by
NSF Grant  PHY-97-22072  and L.R. by D.O.E. cooperative agreement
DE-FC02-94ER40818 and by INFN `Bruno Rossi' Fellowship.


\begin{thebibliography}{ll}
\bibitem{maldacena}J. Maldacena, `The Large $N$ Limit of Superconformal
Theories and Supergravity', Adv.Theor.Math.Phys. {\bf 2} (1998) 231-252, 
hep--th/9711200.
\bibitem{polyakov}{S.S. Gubser, I.R. Klebanov and A.M. Polyakov,
`Gauge Theory Correlators from Non--critical String Theory', Phys.Lett. 
{\bf B428}  (1998) 105-114,
hep--th/9802109.}
\bibitem{witten}{E. Witten, `Anti--de Sitter
Space and Holography', Adv.Theor.Math.Phys. {\bf 2} (1998) 253-291,  hep--th/9802150.}


\bibitem{canadians}  W. Muck, K. S. Viswanathan, `Conformal Field Theory 
Correlators from Classical Scalar Field Theory on $AdS_{d+1}$',
 Phys.Rev. D58 (1998) 041901, hep--th/9804035.
\bibitem{august}
D.Z.~Freedman, S.D.~Mathur, A.~Matusis and L.~Rastelli,
``Comments on 4 point functions in the CFT/AdS correspondence,"
Phys. Lett. {\bf B452}, 61 (1999),
hep-th/9808006.
\bibitem{dhfgauge} E.~D'Hoker and D.Z.~Freedman,
``Gauge boson exchange in $AdS_{d+1}$,"
Nucl. Phys. {\bf B544}, 612 (1999)
hep-th/9809179.
\bibitem{liu}H. Liu, `Scattering in Anti--de Sitter space and operator
product expansion', hep--th/9811152.
\bibitem{dhfscalar}E. D'Hoker and D. Z. Freedman, `General scalar
exchange in $AdS_{d+1}$', hep--th/9811257, to appear in Nucl. Phys. {\bf B}.

\bibitem{march99}
E.~D'Hoker, D.Z. Freedman, S.D. Mathur, A. Matusis, L. Rastelli,
``Graviton exchange and complete 4--point functions in the AdS/CFT
correspondence,'' hep-th/9903196.

\bibitem{chsch}
G.~Chalmers and K.~Schalm,
``The Large $N_c$ limit of four point functions in ${\cal N}=4$
 superYang-Mills theory
                  from Anti-de Sitter supergravity,"
hep-th/9810051.
``Holographic normal ordering and multiparticle states in the AdS / CFT
                  correspondence," hep-th/9901144.


\bibitem{vannieuv}{H. J. Kim, L. J. Romans, and P. van Nieuwenhuizen,
`The Mass Spectrum Of Chiral $N=2$ $D=10$ Supergravity on $S^5$',
Phys. Rev. {\bf D32} (1985) 389.}


\bibitem{compactifications}S.~Minwalla,
``Particles on AdS(4/7) and primary operators on M(2)-brane and M(5)-brane
                  world volumes,"
JHEP {\bf 10}, 002 (1998)
hep-th/9803053.\\
O.~Aharony, Y.~Oz and Z.~Yin,
``M theory on AdS(p) x S(11-p) and superconformal field theories,"
Phys. Lett. {\bf B430}, 87 (1998)
hep-th/9803051.\\
S.~Deger, A.~Kaya, E.~Sezgin and P.~Sundell,
``Spectrum of D = 6, N=4b supergravity on AdS in three-dimensions x S3,"
Nucl. Phys. {\bf B536}, 110 (1998)
hep-th/9804166.\\
J.~de Boer,
``Six-dimensional supergravity on S(3) x AdS(3) and 2-D conformal field
                  theory,"
hep-th/9806104.

\bibitem{prop} C. Fronsdal, Phys. Rev {\bf D10} (1974) 589;\\
C.P. Burgess  and C. A. Lutken,` Propagators and Effective Potentials in
Anti-de Sitter Space', Nucl. Phys. {\bf B272} (1986) 661;\\
T.  Inami and H. Ooguri, `One Loop Effective Potential in Anti-de Sitter 
Space', Prog. Theo. Phys. {\bf 73} (1985) 1051;\\
C.J.C. Burges, D.Z. Freedman, S.Davis, and G.W. Gibbons, `Supersymmetry
in Anti-de Sitter Space', Ann. Phys. {\bf 167} (1986) 285. 

\bibitem{bfbound}P.~Breitenlohner and D.Z.~Freedman,
``Positive Energy In Anti-De Sitter Backgrounds And Gauged Extended
                  Supergravity,"Phys. Lett. {\bf 115B}, 197 (1982);
``Stability In Gauged Extended Supergravity,"
Ann. Phys. {\bf 144}, 249 (1982).\\
L. Mezincescu and P.K. Townsend, `Stability at a Local Maximum
in Higher Dimensional anti--de Sitter Space and Applications to
Supergravity', Ann. Phys. {\bf 160} (1985) 406.




\bibitem{april}D.Z. Freedman,  S. D. Mathur, A. Matusis, and L. Rastelli,
`Correlations functions in the AdS/CFT correspondence', 
Nucl. Phys. {\bf B546}, 96 (1999), hep--th/984058.



\bibitem{slanski}R. Slansky, `Group theory for unified model building'.
Phys. Rep. {\bf 79} (1981) 1


\bibitem{bateman}A. Erdelyi, {\it Bateman Manuscript Project, Higher
Transcendental Functions}, Krieger Publ. Comp. (1981) Vol. I

\bibitem{january}
E.~D'Hoker, D.Z. Freedman, S.D. Mathur, A. Matusis, L. Rastelli,
``Graviton and gauge boson propagators in $AdS_{d+1}$,"
hep-th/9902042.

\bibitem{allenjacobson}B. Allen and T. Jacobson, `Vector Two Point Functions in
Maximally Symmetric Spaces', Commun. Math. Phys.{\bf 103} (1986) 669. 










\end{thebibliography}
\end{document}